\documentclass[
  10pt,
  preprint,
  amsmath,amssymb,
  aps
]{revtex4-2}

\usepackage[utf8]{inputenc}
\usepackage{fontspec}
\usepackage[english]{babel}

\usepackage{mathtools}
\usepackage{bm}
\usepackage{mathdots}
\usepackage{amsthm}  
\usepackage{ytableau}

\usepackage{graphicx}
\usepackage{dcolumn}

\usepackage{xcolor}   
\usepackage{braket}
\usepackage{booktabs}
\usepackage{nicefrac}
\AtBeginEnvironment{thebibliography}{\footnotesize}

\theoremstyle{definition}

\theoremstyle{plain}

\theoremstyle{remark}

\usepackage{hyperref}
\hypersetup{hidelinks}
\usepackage{glossaries}
\ytableausetup{smalltableaux}

\makeglossaries
\newacronym{agf}{AGF}{average gate fidelity}
\newacronym{cptp}{CPTP}{completely positive and trace preserving}
\newacronym{cg}{CG}{Clebsch--Gordan}
\newacronym{spam}{SPAM}{state preparation and measurement}
\newacronym{wot}{WOT}{wonderful orthogonality theorem}
\newacronym{rb}{RB}{randomized benchmarking}
\newacronym{gt}{GT}{Gelfand-Tsetlin}

\begin{document}

\def\bibfont{\scriptsize}

\title{GroupFunctions.jl: computing individual entries of the
irreducible representations of the unitary group U(d)}

\author{David Amaro-Alcalá}
\thanks{\href{https://orcid.org/0000-0001-8137-2161}{ORCID: 0000-0001-8137-2161}}
\author{Konrad Szymański}
\thanks{\href{https://orcid.org/0000-0001-7676-1605}{ORCID: 0000-0001-7676-1605}}
\affiliation{Research Centre for Quantum Information, Institute of
Physics, Slovak Academy of Sciences, Dúbravská cesta 9, Bratislava,
Slovakia}

\date{2026-07-07}

\begin{abstract}
\href{https://github.com/davidamaro/GroupFunctions.jl}{\texttt{GroupFunctions.jl}}\footnote{David
  Amaro-Alcalá developed the package, its algorithms, and its test
  suite, and prepared the initial documentation. Konrad Szymański
  substantially revised and expanded the documentation and developed
  additional examples, and contributed to the standardisation of the
  API.} is a Julia library for computing individual matrix elements of
irreducible representations of \(\mathrm{U}(d)\). These matrix elements,
called group functions, can be evaluated symbolically or numerically.
For \(\mathrm{SU}(2)\), they reduce to the Wigner \(D\)-functions. The
library computes these matrix elements in a carrier-space basis
enumerated by Gelfand-Tsetlin patterns. It
can also compute entire representation operators, construct input
unitaries from parameterisations common in quantum optics, translate
Gelfand--Tsetlin patterns into occupation-number kets, and compute the
associated Schur functions. Results can be exported in a form compatible
with Mathematica.
\end{abstract}

\maketitle

\section{Statement of need}\label{statement-of-need}

Representations of the unitary group \(\mathrm{U}(d)\) arise in many
subfields of physics and mathematics, and computations often reduce to
evaluating their matrix elements, called group functions. For an irrep
labelled by \(\lambda\), the corresponding object is

\[ D^{(\lambda)}_{\mathrm{out},\mathrm{init}}(U) = \langle \mathrm{out} \mid D^{(\lambda)}(U) \mid \mathrm{init} \rangle ,\]

where \(\mathrm{init}\) and \(\mathrm{out}\) denote basis states in the
representation carrier space.

In mathematics, summing the diagonal group functions gives the trace of
the representation matrix of \(U\). This trace is the character of the
representation and the Schur polynomial of the eigenvalues of \(U\), an
object central to algebraic combinatorics and symmetric function theory.

In quantum physics, group functions appear in several settings. In
quantum optics, a group function gives the transition amplitude of
photons through a linear optical network. The same object can be used to
characterise quantum devices~\citep{xmlq-vssg} and to describe the
symmetry properties of states~\citep{OttoSzymanski2024}. One important
subproblem is boson sampling~\citep{Aaronson2011}, where the transition
amplitude reduces to a permanent whose evaluation is classically
computationally hard. Whether noisy real-world quantum devices can
perform this computation remains an active question.

Some of these tasks require only a numerical estimate, whereas others
require an exact symbolic group function. \texttt{GroupFunctions.jl}
addresses the latter need. Although related packages exist (see below),
to our knowledge, none is designed to compute individual
representation-matrix entries symbolically. When applicable, computing
an individual entry is more efficient than assembling the entire matrix.

\subsection{Similar software}\label{similar-software}

Several existing packages relate to \texttt{GroupFunctions.jl} but
address different problems. \texttt{SUNRepresentations.jl}~\citep{SUNRepresentations}
and the algorithm of~\citet{Alex2011} (with
appendix code) compute \(\mathrm{SU}(d)\) Clebsch-Gordan coefficients.
\texttt{RepLAB}~\citep{RepLAB} supports manipulating irreducible
representations of various groups, including \(\mathrm{U}(d)\), but
provides only indirect numerical access to group functions.
\texttt{IntegrateUnitary.jl}~\citep{IntegrateUnitary} performs symbolic
integration over compact groups rather than evaluating representation
matrices. \texttt{haarpy}~\citep{cardin2024haarpy} implements
Weingarten-calculus methods. Other libraries focus on quantum optics.
\texttt{BosonSampling.jl}~\citep{Seron2024}, \texttt{Perceval}~\citep{Heurtel2023}, and \texttt{QOptCraft}~\citep{QOptCraft}
numerically model linear optical devices, while
\texttt{The\ Walrus}~\citep{Gupt2019} helps compute amplitudes for Gaussian boson sampling.
These packages do not target the symbolic computation of individual
representation-matrix entries, which is the primary purpose of
\texttt{GroupFunctions.jl}.

\section{Example use and
documentation}\label{example-use-and-documentation}

The library primarily computes matrix elements of a representation
between basis states, with auxiliary functions that support calculations
in quantum optics. These matrix elements can be computed from a symbolic
matrix.
\begin{verbatim}
julia> U = su2_block_symbolic(2,1);

julia> U
2×2 Matrix{SymEngine.Basic}:
 v_1_1  v_1_2
 v_2_1  v_2_2

julia> group_function([2,0], U)[1]
3×3 Matrix{SymEngine.Basic}:
             v_2_2^2        sqrt(2)*v_2_2*v_2_1              v_2_1^2
 sqrt(2)*v_1_2*v_2_2  v_1_1*v_2_2 + v_1_2*v_2_1  sqrt(2)*v_1_1*v_2_1
             v_1_2^2        sqrt(2)*v_1_1*v_1_2              v_1_1^2
\end{verbatim}

The following example evaluates a matrix element between states
of the \(\mathrm{U}(10)\) symmetric irrep corresponding to \(7\) bosons.

\begin{verbatim}
lambda = [7,0,0,0,0,0,0,0,0,0];
basis=basis_states(lambda); # integer partition and basis
init = findfirst(gt -> occupation_number(gt)==[0,0,0,1,1,1,1,1,1,1],basis); 
out  = findfirst(gt -> occupation_number(gt)==[1,1,1,1,1,1,1,0,0,0],basis);
group_function(lambda, basis[init],basis[out]) # symbolic matrix element
\end{verbatim}

\texttt{GroupFunctions.jl} is currently single-threaded. On an AMD Ryzen
7 PRO 4750U laptop CPU, the symbolic computation in the example took
approximately 4 seconds after compilation warmup, without reusing cached
intermediate or final results.

This functionality supports more complex calculations. For example, the
library has been used to numerically evaluate \(\mathrm{SU}(d)\) group
characters in randomised benchmarking~\citep{xmlq-vssg}. Further
examples, applications, and mathematical background are available in
\href{https://davidamaro.github.io/GroupFunctions.jl/dev/}{the
documentation}.

\section{Design choices}\label{design-choices}

This library provides a unified method for computing representation
matrix elements of \(\mathrm{U}(d)\) irreps specified by integer
partitions of length at most \(d\), including computations with symbolic
input matrices. Several computational routes are possible in principle.
For symmetric irreps, which model fully indistinguishable bosons, one
can manipulate states as polynomials of creation operators applied to
the vacuum. Another approach constructs and exponentiates the Lie
algebra generators in the chosen representation. Both approaches are
computationally expensive. More restricted methods based on generating
functions also exist~\citep{prakash1996wigner}.

The author of the original package chose the Grabmeier-Kerber formula~\citep{Grabmeier1987}
as the most general solution. It expresses the
matrix element as a sum of monomials in the entries of the input matrix,
weighted by the irreducible representation and the states in question.
Our implementation optimises the enumeration of the double cosets that
index this sum by grouping permutations that contribute the same
monomial. This implementation provides the library's main function,
\texttt{group\_function}.

The computational cost of a call to \texttt{group\_function} is controlled by
three related representation-theoretic quantities. The length
$m=\operatorname{length}(\lambda)$ fixes the group \(\mathrm{SU}(m)\), while
$d_\lambda(m)=\operatorname{length}(\texttt{basis\_states}(\lambda))$ is the
dimension of its Gelfand--Tsetlin representation space. Increasing $m$ at
fixed Young shape means appending trailing zeros to $\lambda$; this generally
increases $d_\lambda(m)$, but our benchmarks show only moderate growth for a
single matrix element. A more substantial increase
occurs when the number of boxes $N=|\lambda|=\sum_i\lambda_i$ grows or when
$\lambda$ labels a high-dimensional $S_N$ irrep. Its dimension $f^\lambda$
equals the number of standard Young tableaux of shape $\lambda$, and larger
values require more representation-matrix work. This is
consistent qualitatively with Bürgisser's analysis of evaluating
general-linear-group representations in Gelfand--Tsetlin bases~\citep{Buergisser2000}.

For this implementation, the symmetric-group enumeration adds another
bottleneck: \texttt{group\_function} constructs double cosets and explicitly
iterates over their permutation elements. Together, these double cosets
partition $S_N$, so the underlying work may involve as many as $N!$
permutations, with Young-representation calculations performed for them.
Thus, a more faithful mental model is
\begin{equation}
\operatorname{Cost}\bigl(\texttt{group\_function}(\lambda,i,j)\bigr)
\approx
\operatorname{poly}\!\bigl(d_\lambda(m),f^\lambda\bigr)
\times \text{permutation/double-coset work},
\end{equation}
where the last factor can approach $N!$ in the current algorithm. In
practice, the limit is therefore usually reached by increasing $N$ or by
choosing a Young diagram with large $f^\lambda$, rather than by increasing
$m$ through trailing zeros alone.

Internally, the algorithm represents basis states as semistandard Young
tableaux. The user-facing functions expose the equivalent
Gelfand-Tsetlin patterns~\citep{GelfandTsetlin1950} through the
\texttt{GTPattern} data structure
and provide utility functions for common quantum optics scenarios, such
as \texttt{occupation\_number}.

\section{Availability}\label{availability}

The library is available under the MIT licence and can be installed
through the Julia package manager:

\begin{verbatim}
 ] add GroupFunctions
\end{verbatim}

\section{AI usage disclosure}\label{ai-usage-disclosure}

OpenAI Codex (GPT-5.3) assisted with optimising the performance of the
double-coset enumeration at a late stage. The authors developed the
mathematical design and proof of correctness of the optimised algorithm.
Anthropic Claude (Opus 4.8) assisted with code review and language
review of the documentation and manuscript. The authors reviewed and
edited all AI-assisted changes.

\section{Acknowledgements}\label{acknowledgements}

We thank Dr.~Hubert de Guise for helpful discussions and suggestions on
the bibliography, and Dr.~Alonso Botero for suggestions that improved
the presentation. Mitacs CALAREO, DeQHOST APVV-22-0570, QUAS VEGA
2/0164/25, Postdokgrant APD0161, and the Stefan Schwarz programme
supported this work. David Amaro-Alcalá acknowledges the indirect
support of the Government of Alberta and NSERC during his PhD studies at
the University of Calgary.

\bibliographystyle{apsrev4-2}
\bibliography{references.bib}

\end{document}